# Triggerless data acquisition in asynchronous optical-sampling terahertz time-domain spectroscopy based on dual-comb system

Makoto Okano[1,2,†] and Shinichi Watanabe[2,‡]

[1]*Department of Communications Engineering, National Defense Academy, 1-10-20, Hashirimizu, Yokosuka, Kanagawa 239-8686, Japan*

[2]*Department of Physics, Faculty of Science and Technology, Keio University, 3-14-1 Hiyoshi, Kohoku-ku, Yokohama, Kanagawa 223-8522, Japan*

[†]*makokano@nda.ac.jp*

[‡]*watanabe@phys.keio.ac.jp*

## Abstract

By using two mutually phase-locked optical frequency combs with slightly different repetition rates, we demonstrate asynchronous optical-sampling terahertz time-domain spectroscopy (ASOPS THz-TDS) without using any trigger signals or optical delay lines. Due to a tight stabilization of the repetition frequencies, it was possible to accumulate the data over 48 minutes in a triggerless manner without signal degradation. The fractional frequency stability of the measured terahertz signal is evaluated to be $\sim 8.0 \times 10^{-17}$ after 730 s. The frequency accuracy of the obtained terahertz spectrum is ensured by phase-locking the two frequency combs to a frequency standard. To clarify the performance of our system, we characterized the absorption line of water vapor around 0.557 THz. The good agreement of the measured center frequency and linewidth with the values predicted from the HITRAN database verifies the suitability of our ASOPS THz-TDS system for precise measurements.

# 1. Introduction

During the past decades, terahertz spectroscopy has become a technique of great interest in fundamental science and development of various applications, because of useful features such as excitation with low-energy photons, and a usually high transmissivity of materials in this frequency range [1]. In particular, terahertz time-domain spectroscopy (THz-TDS) allows us to simultaneously measure the phase and amplitude of terahertz light. This remarkable feature allows us to directly determine the complex refractive index or complex dielectric function of a material [2]. Thus, THz-TDS has been widely used to investigate the optical and electrical properties of various materials. In addition, THz-TDS can also be used to study molecular gases, because most vibration modes of molecules are in the terahertz frequency range [3,4]. In conventional THz-TDS, a single light source is used and the spectral information for a given sample is obtained by the Fourier transform of the time-domain waveform of the terahertz pulse reflected or transmitted by the sample. Therefore, the spectral resolution of the conventional approach is restricted by the measurable time range, which is in general determined by the length of the used translation stage of the optical delay line. For a high spectral resolution, it is thus essential to employ a long stage, but this causes difficulties in the optical alignment and results in long measurement times.

To obtain a high spectral resolution without the restrictions imposed by a long optical delay line, various approaches have been proposed and demonstrated. The approaches based on asynchronous optical sampling (ASOPS) [5–12] and electronically controlled optical sampling (ECOPS) [13,14] are promising candidates for THz-TDS with enhanced spectral resolution. In these methods, two optical pulses with different repetition frequencies are used for generation and detection of the terahertz pulse. The time interval between the optical pulse for terahertz pulse generation and that for the detection of the terahertz time domain waveform sweeps automatically due to the difference in the repetition frequencies, and thus the time-domain waveform can be obtained without using any optical delay lines. Since the measurable time range is typically on the order of tens of nanoseconds (determined by the inverse of the repetition frequency of the pulse for detection), it is significantly longer than that of conventional THz-TDS systems with optical delay lines.

Regarding the generation of two optical pulses with different repetition frequencies, two typical methods can be found in the literature: One is the usage of two pulsed lasers with different repetition frequencies [16, 17], and the other is the generation of two optical pulses with different repetition frequencies by a single pulsed laser using optical modulation of the output pulses [18] or by using a bidirectional mode-locked fiber laser [19, 20]. To realize ASOPS and ECOPS THz-TDS, a complete stabilization of the repetition frequencies of the two optical pulses is essential. Recently, an adaptive sampling method, which is an alternative technique without complete stabilization, has also been proposed [21]. In either case, an electrical trigger signal is usually used to precisely accumulate the time-resolved data to improve the signal-to-noise ratio. However, the limited rise time of the detector that generates the trigger pulse used for timing the data acquisition may lead to a timing jitter, which can cause a reduction in the signal intensity of the averaged terahertz waveform [9]. To eliminate the effect of the timing jitter of such a trigger system, the development of measurement system in which no trigger signal is required for data acquisition, i.e., a triggerless system, is one of the possible solutions. Recently, a triggerless THz-TDS system based on two pulsed lasers has been reported [22]. However, Vieira et al. pointed out that an accumulation over a long time of 42 min causes a reduction in the signal intensity due to the timing jitter of the used light sources. Thus, to realize a more accurate triggerless THz-TDS system, a more accurate frequency stabilization of the light sources is indispensable.

In this paper, we report on a triggerless ASOPS THz-TDS system that uses two mutually phase-locked optical frequency combs (OFCs) as light sources. In dual-comb spectroscopy (DCS) with fully stabilized OFCs [23], the repetition frequencies of the two OFCs are stable over a long period of time. This feature is important in reducing the timing jitter between the optical pulses from the OFCs, resulting in the ability to accumulate data without a degradation of the signal intensity even if no trigger signal is used for timing the data acquisition. Thus, we developed a triggerless DCS-based ASOPS THz-TDS system with a high frequency resolution and frequency information that is traceable to a global-positioning-system-controlled rubidium (Rb) clock. By performing a measurement with an accumulation time of about 48 min, we show that no signal degradation occurs during the accumulation of 16160 single waveform scans, which indicates a high stability of this system. Additionally, we measure the absorption line of the pure rotational transition $1_{10}$–$1_{01}$ of $H_2^{16}O$ to verify the suitability of our system for precise spectroscopic measurements.

## 2. Experimental setup

Our DCS system is composed of two erbium-doped fiber (EDF)-based OFCs with slightly different repetition rates [24]. The OFC for terahertz pulsed generation is referred to as comb1, and the other is referred to as comb2. We used the following procedure to stabilize the two OFCs: The carrier-envelope offset frequency of comb1 ($f_{ceo1}$) and the repetition rate of comb1 ($f_{rep1}$) are phase-locked to RF signals from two different function generators (FGs). The beat frequency between one of the teeth of comb1 and a continuous wave (CW) external cavity laser with a wavelength of ≈1550.11 nm, $f_{beat1}$, is phase-locked to an RF signal. The beat signal between one of the comb teeth of comb2 and the CW external cavity laser, $f_{beat2}$, is also phase-locked to an RF signal from a third FG, and thus comb1 and comb2 are mutually phase-locked in the optical frequency region. The carrier-envelope offset frequency of comb2 ($f_{ceo2}$) is phase-locked to an RF signal from the third FG. As a result of the frequency stabilization of $f_{ceo1}$, $f_{beat1}$, $f_{rep1}$, $f_{ceo2}$, and $f_{beat2}$, the repetition rate of comb2 ($f_{rep2}$) is indirectly stabilized. We set $M \equiv f_{rep1}/(f_{rep2}-f_{rep1})$ to an integer value to achieve the coherent averaging condition [25, 26]. All FGs were referenced to a global-positioning-system-controlled Rb clock. More details of our DCS system are described in [24].

As shown in Fig. 1, the output of each OFC was divided several times. First, a part of the output of each OFC was used to detect $f_{rep}$. Another part was used for the detection of $f_{ceo}$ and $f_{beat}$. The remaining light was divided into two branches for measurements in the terahertz and near-infrared regions. The light of the terahertz branch of comb1 was introduced to a fiber-coupled photoconductive antenna (PCA) for the generation of the terahertz wave, and that of comb2 was introduced to a fiber-coupled PCA for the detection of the terahertz wave. The terahertz wave emitted from the terahertz transmitter (THz-P-Tx, Fraunhofer HHI) was collected by an off-axis parabolic mirror. Then it passed through three off-axis parabolic mirrors and was finally focused on the terahertz receiver (EK-000782, TOPTICA Photonics AG), where the electric-field component of terahertz wave was converted to a photocurrent signal using the output of comb2. To avoid the influence of absorption by water vapor in the ambient atmosphere, we enclosed the four off-axis parabolic mirrors and the two PCAs by a box and purged the box with dried air. The photocurrent was amplified by a wideband current amplifier (SA-605F2, NF Corp.) and sampled by a digitizer (M2p.5962-x4, Spectrum) with a sampling frequency of $f_{rep1}$, where the measurable time range is equal to $1/f_{rep2}$. Because we set $M$ to an integer value, single scans of the terahertz waveform are repeatedly recorded with a period of $M$ sampling points.

The light of the near-infrared branch of each OFC was used to obtain an interference signal between comb1 and comb2 with a fixed-gain balanced amplified photodetector (PDB415C, Thorlabs). The time trace of the interference signal serves as a reference and is hereafter referred to as interferogram (IGM).

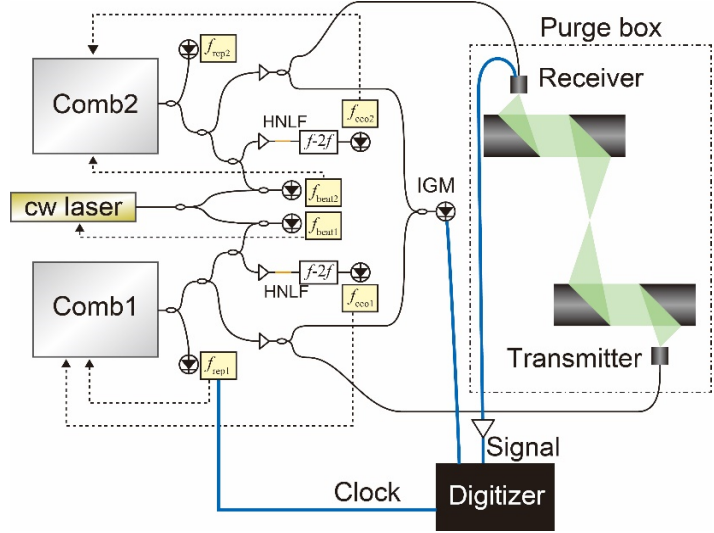

**Fig. 1**. Schematic view of the triggerless DCS-based ASOPS THz-TDS system. The black curves depict optical fibers and the blue curves depict electric cables. The dotted arrows correspond to the feedback for frequency stabilization. The dashed–dotted line represents the purge box. HNLF: highly nonlinear fiber, $f$–$2f$: $f$–$2f$ interferometer with a periodically poled lithium niobate crystal.

### 3. Results and discussion

*3.1 Stability of the triggerless DCS-based ASOPS THz-TDS system*

Figure 2(a) shows the terahertz time-domain waveform obtained by averaging the data of all 16160 single scans recorded during about 48 min (red curve) and compares it with the waveform data of a single scan (black curve). We set $f_{rep1}$ to 61530670 Hz and $M$ to 11028978, which implies $\Delta f_{rep} = f_{rep2} - f_{rep1} = f_{rep1}/M \approx 5.579$ Hz. The temporal window is $\approx 16.3$ ns and the temporal resolution is $\approx 1.47$ fs. The inset of Fig. 2(a) shows a magnified view around a time origin where the peak electric-field of the terahertz time-domain waveform is observed. The averaged waveform exhibits no signal degradation with respect to the single-scan waveform although we did not use any trigger signals during the data acquisition. This indicates that the timing jitter between the two laser pulses is negligibly small.

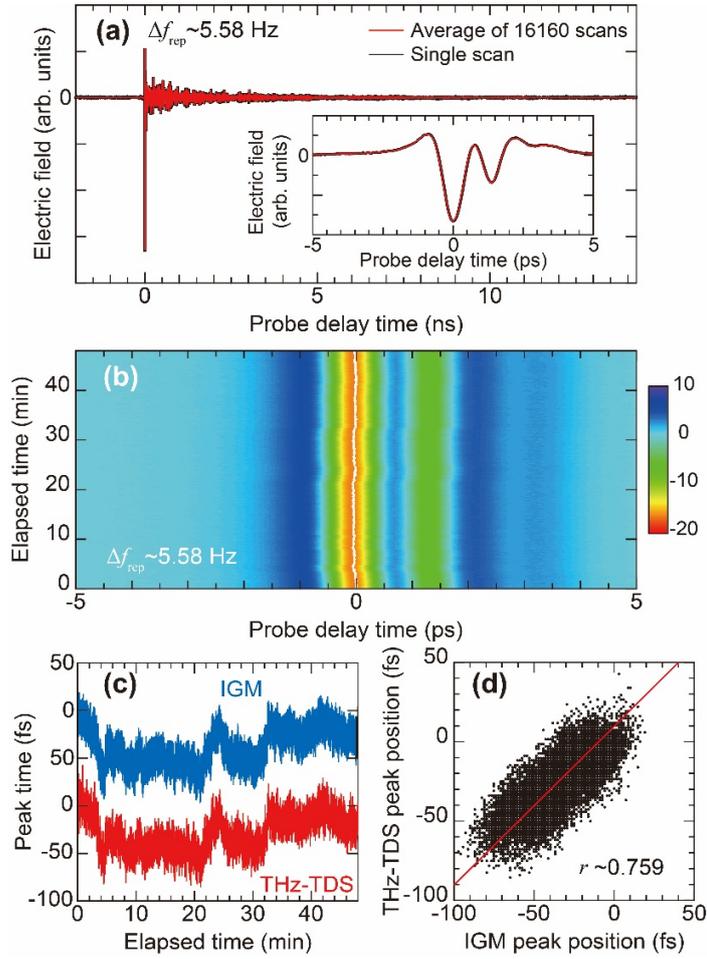

**Fig. 2**. (a) Terahertz time-domain waveform obtained by a single scan (black curve) and that obtained by averaging 16160 single scans (red curve). Inset: Enlarged view in the vicinity of the time origin. (b) Dependence of the measured waveform on the time elapsed since the start of the data accumulation. The white curve corresponds to the temporal position of the peak electric field in each single scan. (c) Dependence of the peak position on the elapsed time for the THz-TDS waveform data [red curve, same data as the white curve in (b)] and that for the IGM data (blue curve). The blue curve is offset for clarity. (d) The correlation between the THz-TDS and IGM peak positions in (c) plotted for all 16160 scans. The red line is a guide to the eye.

To evaluate the stability of our measurement system in the time domain, we investigated the fluctuation of the temporal position of the peak electric-field during the acquisition of the 16160 time-domain waveforms. Figure 2(b) shows a two-dimensional plot of the signal intensity as functions of the probe delay time around the time origin (horizontal axis) and the time elapsed since the start of the data accumulation (vertical axis). The white curve shows the peak positions in the 16160 time-domain waveforms and evidences that the peak position fluctuated slightly during the measurement. The red curve in Fig. 2(c) plots these peak positions using a smaller scale: A fluctuation on the order of ~50 fs can be confirmed. Since this fluctuation is much shorter than the typical period of a terahertz wave (~1 ps), there is almost no signal degradation in the case of averaging as shown in Fig. 2(a).

To gain more insights into the origin of the temporal fluctuation of the THz-TDS peak position [Fig. 2(c); red curve], we compare the individual peak positions with those of the simultaneously measured IGMs [Fig. 2(c); blue curve]. The peak position in each IGM was defined as the peak position of the envelope function obtained by using the Hilbert transform. Both time traces show similar tendencies with respect to the elapsed time. The correlation between the THz-TDS and IGM peak positions is plotted in Fig. 2(d). The Pearson's correlation coefficient is evaluated to be ~0.759, which indicates a strong correlation. This suggests that the fluctuation of the THz-TDS peak position is governed by the stability of the common OFC part consisting of oscillator and amplifier, and not by that of the optical system consisting of the off-axis parabolic mirrors and the PCAs. Both time traces in Fig. 2(c) consist of a fast fluctuation component (fluctuation magnitude on the order of ~50 fs) and several slow large shifts that take place during a few minutes (for instance, around 0–4 min). We attribute the fast fluctuation to detector noise, shot-noise, and laser relative intensity noise [28], and the slow large shifts to environmental changes such as temperature changes. Therefore, the usage of low-noise detectors and laser drivers, and the control of the environment may improve the stability further.

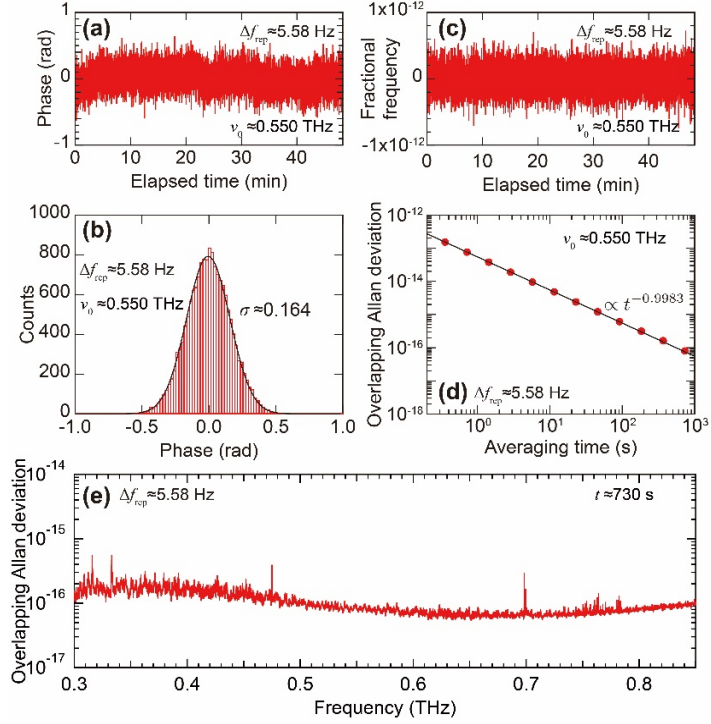

**Fig. 3**. (a) Dependence of the phase fluctuation at ≈0.550 THz on the elapsed time. (b) The histogram of the data shown in (a). The black curve is a fit using a Gaussian function. (c) The fractional frequency $y(t)$ plotted as a function of the elapsed time. (d) Overlapping Allan deviation $\sigma_y(\tau)$ at ≈0.550 THz as a function of the averaging time and (e) the frequency dependence of $\sigma_y(730\ s)$.

To further clarify the details of the stability of our system, we also evaluated the stability of the terahertz signal in the frequency domain. First, we calculated the Fourier transform of each recorded terahertz time-domain waveform. Then, the temporal change of the phase, $\phi(t)$, was evaluated at a certain frequency $v_0$ corresponding to a specific frequency-comb tooth. Figure 3(a) shows $\phi(t)$ at $v_0 \approx 0.550$ THz, where the signal intensity is sufficiently large (a detailed discussion of the frequency bandwidth of our system is provided in the Appendix). The corresponding histogram of $\phi$ is plotted in Fig. 3(b). The histogram shows a normal distribution with a standard deviation of ≈0.164. The phase fluctuations of the terahertz comb teeth in our experiment occurred randomly, and the accuracy of the measured phase can be improved by accumulating the signal. In the case of averaging over all 16160 single scans, the accuracy of the averaged phase is estimated to be ~1.3 mrad, which is remarkably small.

In order to assess the frequency stability of our system in the case of long acquisition times, we calculated the overlapping Allan deviation $\sigma_y(\tau)$, where $\tau$ corresponds to the averaging time. According to [29], $\sigma_y(\tau)$ is based on the fractional frequency $y(t)$, and we derived $\sigma_y(\tau)$ from $\phi(t)$ at $v_0$ as follows:

$$y(t) = \frac{d\phi(t)}{dt} \cdot \frac{1}{2\pi v_0}, \quad (1)$$

$$\sigma_y(\tau) = \sqrt{\frac{1}{2m^2(M-2m+1)} \sum_{j=1}^{M-2m+1} \left\{ \sum_{i=j}^{j+m-1} [y_{i+m} - y_i] \right\}^2}. \quad (2)$$

Here, $M$ is the total number of data points of $y(t)$, and $\tau = m\tau_0$, where $\tau_0$ is the time interval between the data points of $y(t)$ and $m$ is an integer. $y(t)$ is provided in Fig. 3(c).

Figure 3(d) shows $\sigma_y(\tau)$ at $v_0 \approx 0.550$ THz as a function of $\tau$. We obtain $\sigma_y \approx 3.8 \times 10^{-14}$ for an averaging time of 1 s, and $\sigma_y$ reaches about $8.0 \times 10^{-17}$ for an averaging time of 730 s. These values are comparable with the previously reported values in the terahertz frequency range for a system with a CW terahertz light source [30]. We find that $\sigma_y(\tau)$ at $v_0 \approx 0.550$ THz is proportional to $\tau^{-0.9983}$ in the entire time range. The exponent of the power law is almost unity, indicating that white or flicker phase-modulation noise is the main origin of the frequency instability in our system. Figure 3(e) shows the frequency dependence of $\sigma_y$ at $\tau = 730$ s. The observed weak frequency dependence is attributed to the frequency dependence of the amplitude shown in Fig. 4 (a). The data clarifies that $\sigma_y(730\text{ s}) < 5 \times 10^{-16}$ in the entire frequency region. Because the OFCs possess a high frequency stability and accuracy for each comb, our result of a high frequency stability in the terahertz frequency range originates from the OFC technology.

*3.2 Measurement of the absorption spectrum of water vapor*

To clarify the suitability of our system for highly precise spectroscopic measurements, we measured the absorption spectrum of water vapor. In this experiment, we stabilized $f_{rep1}$ to 61530990 Hz and $\Delta f_{rep}$ to ~5.57 Hz. The temperature and the humidity during this experiment were 20.8±0.5°C and 39±5%, respectively (measured by a thermo-hygrometer). To obtain the signal without water vapor, we purged the box with dry air. The acquisition times for both measurements were ≈10 min.

Figure 4(a) shows the amplitude spectra with and without water vapor. Each spectrum was obtained by the Fourier transform of the corresponding averaged time-domain waveform (averaged over 3500 single scans). In the red curve, absorption lines due to water vapor are clearly observed in the range from 0.5 to 1.8 THz. The inset of Fig. 4(a) shows the absorption line of the pure rotational transition $1_{10}$–$1_{01}$ of $H_2^{16}O$, which appears at around 0.557 THz. The absorption spectrum of the $1_{10}$–$1_{01}$ transition of $H_2^{16}O$ calculated from the spectrum in Fig. 4(a) is plotted in Fig. 4(b). By fitting the data to a Lorentzian function, we evaluated a center frequency of 557.130(20) GHz and a half width at half maximum (HWHM) of 3.109(29) GHz for this absorption line.

To compare these values with the data from the HITRAN database, we need to evaluate the impact of air, which acts as the buffer gas in this experiment. According to the previous studies [31–33], the shift coefficient of air ($\Delta v_{air}$) and the broadening coefficient of air ($\gamma_{air}$) are a linear combination of the corresponding coefficients of the constituent gases such as $N_2$ and $O_2$. Here, we assume an air pressure of 760 Torr and that the air consists of 79% $N_2$ and 21% $O_2$. In this case, $\Delta v_{air}$ and $\gamma_{air}$ are given by

$$\Delta v_{air} = 0.79 \Delta v_{N_2} + 0.21 \Delta v_{O_2}, \tag{3}$$

$$\gamma_{air} = 0.79 \gamma_{N_2} + 0.21 \gamma_{O_2}, \tag{4}$$

where $\Delta v_{N_2(O_2)}$ and $\gamma_{N_2(O_2)}$ are the shift and broadening coefficients of $N_2$ ($O_2$), respectively. Using the values reported in Refs. [34, 35], we obtain $\Delta v_{air}$ = 0.250 MHz/Torr and $\gamma_{air}$ = 4.15 MHz/Torr. Based on the HITRAN database (because the absorption of the $1_{10}$–$1_{01}$ transition of $H_2^{16}O$ is located at 556.936 GHz), the center frequency and the HWHM of the $1_{10}$–$1_{01}$ transition at 760 Torr are evaluated to be 557.125(12) GHz and 3.150(12) GHz, respectively. These values can be compared with those derived from the data in Fig. 4; 557.130(20) GHz and 3.109(29) GHz, respectively. The good agreement within the experimental uncertainty suggests a high frequency accuracy of our triggerless DCS-based ASOPS THz-TDS system.

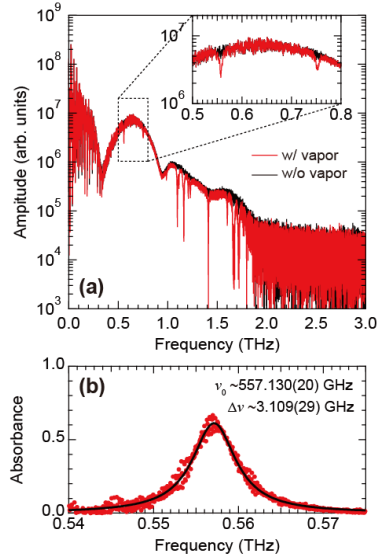

**Fig. 4**. (a) Amplitude spectra with (red curve) and without water vapor (black curve). Inset: Magnified view of the amplitude spectra in the range of 0.5–0.8 THz. (b) Absorption spectrum of water vapor around 0.557 THz derived from (a). The black curve represents the result of fitting the data to a Lorentzian function.

## 4. Conclusions

We have demonstrated a DCS-based ASOPS THz-TDS system that can be operated without using any electric trigger signals. Owing to the usage of a sophisticated DCS system as a light source, the timing jitter between the laser pulses from the two OFCs is very small, which allows us to average the THz-TDS signal over 48 minutes without signal degradation. By comparing the stabilities of the peak positions of the terahertz time-domain waveform and the IGM in the near-infrared region, we found that the stability of the THz-TDS measurement is mainly determined by the stability of the DCS system and the environment. We also assessed the stability of data accumulation over a long time by using the overlapping Allan deviation $\sigma_y$; the fractional frequency stability is $8.0\times10^{-17}$ for an integration time of 730 s at 0.550 THz, and $\sigma_y$(730 s) is $<5\times10^{-16}$ in the range from 0.3 to 0.8 THz. To verify the applicability of our system, the absorption spectrum of water vapor was measured. The measured center frequency and width of the absorption line at around 0.557 THz show a good agreement with the values predicted from the HITRAN database. Thus, we conclude that our system is promising for extremely precise spectroscopic measurements in the terahertz frequency range.

**Appendix: Dependence of the frequency bandwidth of the system on $\Delta f_{rep}$**

In ASOPS THz-TDS systems, the frequency resolution in the terahertz frequency domain is determined by the repetition rate of the pulsed light source ($f_{rep2}$ in our case), and that of the recorded time-domain signal in the RF domain is determined by $\Delta f_{rep}$. If the optical pulse width is sufficiently short to generate a broad terahertz spectrum, the frequency bandwidth of the terahertz pulse is given by $(f_{rep2} \cdot f_{det})/\Delta f_{rep}$, where $f_{det}$ is the frequency bandwidth of the detector. In our case, since the terahertz emitter and receiver have a larger frequency bandwidth than the wideband current amplifier, $f_{det}$ is determined by the bandwidth of the wideband current amplifier (~250 kHz).

Figure 5 shows the measured power spectral density for different values of $\Delta f_{rep}$. We can confirm a narrower frequency bandwidth for larger values of $\Delta f_{rep}$. In the case of $\Delta f_{rep}$=83.9 Hz, $(f_{rep2} \cdot f_{det})/\Delta f_{rep} \approx 0.183$ THz, which agrees with the observed frequency range. On the other hand, in the case of $\Delta f_{rep}$=5.57 Hz, the bandwidth determined by the detector bandwidth is estimated to be ~2.76 THz, while the observed terahertz bandwidth is significantly narrower. These results show that the measurable frequency range in our system is mainly determined by $f_{det}$ in the case of larger $\Delta f_{rep}$ values, while it is mainly determined by the pulse width of our frequency comb laser (~a few hundreds of femtoseconds) in the case of smaller Δ$f_{rep}$ values.

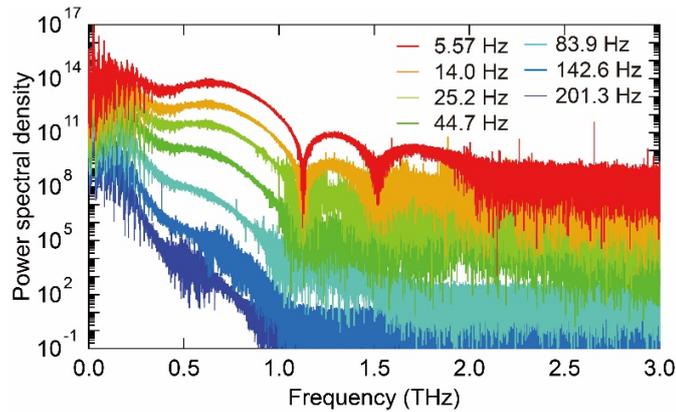

**Fig. 5**. Power spectral density of the measured terahertz time-domain waveform for different values of Δ$f_{rep}$.

**Funding.** This work was partially supported by JSPS KAKENHI Grant Nos. JP18H02040 and JP22K03557, the MEXT Quantum Leap Flagship Program (MEXT Q-LEAP) Grant No. JPMXS0118067246, and JST CREST Grant No. JPMJCR19J4, Japan.